\documentclass[aps,prc,twocolumn,superscriptaddress,floatfix,nofootinbib,longbibliography]{revtex4-1}

\usepackage[utf8]{inputenc}
\usepackage[english]{babel}
\usepackage{amssymb,amsthm,amsmath,amstext,amsbsy,amsopn}
\usepackage{bbm}
\usepackage{graphicx}
\usepackage{isotope}
\usepackage{float}
\usepackage[dvipsnames]{xcolor}
\usepackage{tabularx}

\usepackage[
    colorlinks=true,
    linkcolor=blue,
    citecolor=blue,
    urlcolor=blue
]{hyperref}

\newcommand{\la}{\langle}
\newcommand{\ra}{\rangle}

\newcommand{\lang}{\ensuremath{l}}
\newcommand{\langpr}{\ensuremath{l'}}

\newcommand{\uncoupledchannel}[2]{$^1{#1}_{#2}$}
\newcommand{\coupledchannel}[3]{$^3{#1}_{#3}$-$^3{#2}_{#3}$}

\newcommand{\ai}{{\emph{ab initio}}}

\newcommand{\RSVD}{\ensuremath{R_\text{SVD}}}

\newcommand{\MeV}{\ensuremath{\text{MeV}}}

\newcommand{\fm}{\ensuremath{\text{fm}}}

\newcommand{\NN}{\text{NN}}

\newcommand{\cost}{\text{cost}}
\newcommand{\contr}{\mathcal{C}}
\newcommand{\stor}{\mathcal{M}}
\newcommand{\fcontr}{\mathcal{\tilde C}}
\newcommand{\fstor}{\mathcal{\tilde M}}
\newcommand{\scaling}[1]{\mathcal{O}(#1)}

\newcommand*{\nxlo}[1]{N${}^{#1}$LO}

\newcommand{\tfls}{{SVD-LS}}

\newcommand{\eg}{\textit{e.g.}}
\newcommand{\ie}{\textit{i.e.}}

\newcommand{\LippSchw}{{Lippmann-Schwinger}}

\DeclareMathOperator{\Tr}{Tr}

\begin{document}

\allowdisplaybreaks

\title{Least-square approach for singular value decompositions of scattering problems
}

\author{A.~Tichai}
\email{alexander.tichai@physik.tu-darmstadt.de} 
\affiliation{Technische Universit\"at Darmstadt, Department of Physics, 64289 Darmstadt, Germany}
\affiliation{ExtreMe Matter Institute EMMI, GSI Helmholtzzentrum f\"ur Schwerionenforschung GmbH, 64291 Darmstadt, Germany}
\affiliation{Max-Planck-Institut f\"ur Kernphysik, Saupfercheckweg 1, 69117 Heidelberg, Germany}

\author{P.~Arthuis}
\email{parthuis@theorie.ikp.physik.tu-darmstadt.de}
\affiliation{Technische Universit\"at Darmstadt, Department of Physics, 64289 Darmstadt, Germany}
\affiliation{ExtreMe Matter Institute EMMI, GSI Helmholtzzentrum f\"ur Schwerionenforschung GmbH, 64291 Darmstadt, Germany}

\author{K.~Hebeler}
\email{kai.hebeler@physik.tu-darmstadt.de}
\affiliation{Technische Universit\"at Darmstadt, Department of Physics, 64289 Darmstadt, Germany}
\affiliation{ExtreMe Matter Institute EMMI, GSI Helmholtzzentrum f\"ur Schwerionenforschung GmbH, 64291 Darmstadt, Germany}
\affiliation{Max-Planck-Institut f\"ur Kernphysik, Saupfercheckweg 1, 69117 Heidelberg, Germany}

\author{M.~Heinz}
\email{mheinz@theorie.ikp.physik.tu-darmstadt.de}
\affiliation{Technische Universit\"at Darmstadt, Department of Physics, 64289 Darmstadt, Germany}
\affiliation{ExtreMe Matter Institute EMMI, GSI Helmholtzzentrum f\"ur Schwerionenforschung GmbH, 64291 Darmstadt, Germany}
\affiliation{Max-Planck-Institut f\"ur Kernphysik, Saupfercheckweg 1, 69117 Heidelberg, Germany}

\author{J.~Hoppe}
\email{jhoppe@theorie.ikp.physik.tu-darmstadt.de}
\affiliation{Technische Universit\"at Darmstadt, Department of Physics, 64289 Darmstadt, Germany}
\affiliation{ExtreMe Matter Institute EMMI, GSI Helmholtzzentrum f\"ur Schwerionenforschung GmbH, 64291 Darmstadt, Germany}

\author{A.~Schwenk}
\email{schwenk@physik.tu-darmstadt.de}
\affiliation{Technische Universit\"at Darmstadt, Department of Physics, 64289 Darmstadt, Germany}
\affiliation{ExtreMe Matter Institute EMMI, GSI Helmholtzzentrum f\"ur Schwerionenforschung GmbH, 64291 Darmstadt, Germany}
\affiliation{Max-Planck-Institut f\"ur Kernphysik, Saupfercheckweg 1, 69117 Heidelberg, Germany}

\author{L.~Zurek}
\email{lzurek@theorie.ikp.physik.tu-darmstadt.de}
\affiliation{Technische Universit\"at Darmstadt, Department of Physics, 64289 Darmstadt, Germany}
\affiliation{ExtreMe Matter Institute EMMI, GSI Helmholtzzentrum f\"ur Schwerionenforschung GmbH, 64291 Darmstadt, Germany}

\begin{abstract}
It was recently observed that chiral two-body interactions can be efficiently represented using matrix factorization techniques such as the singular value decomposition.
However, the exploitation of these low-rank structures in a few- or many-body framework is nontrivial and requires reformulations that explicitly utilize the decomposition format.
In this work, we present a general least-square approach that is applicable to different few- and many-body frameworks and allows for an efficient reduction to a low number of singular values in the least-square iteration.
We verify the feasibility of the least-square approach by solving the Lippmann-Schwinger equation in factorized form.
The resulting low-rank approximations of the $T$ matrix are found to fully capture scattering observables.
Potential applications of the least-square approach to other frameworks with the goal of employing tensor factorization techniques are discussed.
\end{abstract}

\maketitle

\section{Introduction}

\textit{Ab initio} calculations of nuclear many-body systems have seen significant progress over the past decade due to computational advances, interactions based on chiral effective field theory (EFT), and developments in the field of quantum many-body theory~\cite{Herg20review,Hebe203NF}.
In particular, chiral two- and three-nucleon interactions~\cite{Epel09RMP,Mach11PR,Hebe11fits,Ekst15sat,Ente17EMn4lo,Epel19nuclfFront,Jian20N2LOGO} not only provide a systematic expansion rooted in QCD but also enable estimates of theoretical uncertainties~\cite{Epel15improved,Furn15uncert,LENPIC2019}.
The combination with systematically improvable many-body methods has led to unprecedented studies targeting heavier and exotic nuclei~\cite{Morr17Tin,Arthuis2020a,Stroberg2021,Miyagi2021,Hu21Pb}.

In this context second-quantized representations of two- and many-body operators provide the fundamental input for all basis-expansion methods, \eg{}, many-body perturbation theory (MBPT)~\cite{Holt14Ca,Tich16HFMBPT,Tichai18BMBPT,Dris17MCshort,Tichai2020review}, coupled cluster (CC) theory~\cite{Hage14RPP,Bind14CCheavy,Hage16CCPS}, self-consistent Green's function (SCGF) theory~\cite{Dick04PPNP,Carb13nm,Soma20SCGF},  and the in-medium similarity renormalization group (IMSRG)~\cite{Herg16PR,Stroberg2019,Heinz2021}.
While the use of a single-particle basis is very convenient in practice, the operator representation in this basis requires an extensive number of basis functions to enable robust extractions of nuclear observables.
As such, finding and employing alternative operator bases provides a promising alternative to more efficiently represent the underlying objects.
Recently, the use of low-rank operator expansions obtained from a truncated singular value decomposition was shown to provide excellent approximations to chiral two-nucleon interactions~\cite{Tich21SVDNN,Zhu21SVD}.
Based on such low-rank approximations it was shown that two-nucleon scattering as well as ground-state properties of medium-mass nuclei and the nuclear-matter energy can be very well reproduced from low-rank approximations of chiral interactions.

One of the major advantages of such low-rank approximations is their potential to reduce the required computational resources with respect to storage and the operation cost associated to tensor contractions such as matrix multiplications.
However, fully exploiting the structure of factorized many-body operators requires a reformulation of the underlying many-body approach in terms of the decomposition components themselves.
While this strategy has been extensively studied in quantum chemistry (see, \eg{}, Refs.~\cite{Kinoshita2003,Hohenstein2013,Schutski2017,Parrish2019,Hohenstein2019,Lesiuk2021,Hohenstein2022}), in nuclear physics we are just starting to explore such ideas in many-body calculations.
Finally, the use of factorized tensor representations is at the heart of the density matrix renormalization group (DMRG), which has been used with great success in condensed matter physics and quantum chemistry~\cite{White1992,Schollwoeck2011,Baiardi2020}.
Recently, the DMRG ansatz has also been employed in nuclear physics applications~\cite{Papenbrock2005,Legeza2015,Foss17TetraN,Tichai2022dmrg}.
In addition, there have been various applications employing factorization techniques in fitting procedures or as diagnostic tools~\cite{Bertsch2005,Bertsch2009,Stoitsov2010,Johnson2010}.

Ultimately, factorization techniques may provide a way of extending the reach of \ai{} nuclear structure calculations to heavier and more exotic systems. 
The computational demands of such calculations are due to i)~the increase in model-space dimension necessary to obtain converged calculations of heavy  nuclei~\cite{Miyagi2021,Hu21Pb}, ii)~the need for refined truncation schemes in the many-body expansion~\cite{Heinz2021,Barbieri2021,Novario2020a}, and iii)~the use of symmetry-unrestricted bases to account for nuclear deformation effects in open-shell nuclei~\cite{Novario2020a,Frosini2021mrIII,Hagen2022PCC,Yuan2022}. 
Exploiting the low-rank properties of nuclear interactions can help to push the present frontiers to access significantly larger many-body spaces and thus better capture correlations.
This strategy is complementary to ongoing efforts to compress calculations using importance truncation methods~\cite{Roth09ImTr,Tichai2019pre,Porr21ITSCGF,HoppeIT2022} and to construct improved bases with superior convergence properties~\cite{Tich19NatNCSM,Novario2020a,HoppeNatOrb2020,Fasano2021nat}.

In this work, we present a novel strategy that builds upon a least-square minimization of the decomposition error of the unknown tensor in a given framework.
The equations we obtain operate exclusively on the decomposed factors without reconstructing the full operators at any point.
Additionally, they are independent of the details of the few- or many-body method.
For this reason, the least-square approach is a general strategy that can be used to reformulate few- and many-body methods to exploit tensor factorization techniques.
As a proof of concept we apply the least-square approach to the Lippmann-Schwinger equation and the full $T$ matrix. 
This paper is organized as follows.
In Sec.~\ref{sec:lse}, the least-square approach is introduced.
Section~\ref{sec:lsls} provides the application to the Lippmann-Schwinger equation including numerical results for low-rank $T$ matrices.
Finally, we conclude with an outlook on future perspectives in Sec.~\ref{sec:outlook}.

\section{Least-square factorization}
\label{sec:lse}

\subsection{General rationale}

In the following, we aim at deriving a factorized form of algebraic equations of the form 
\begin{align}
    T = f(T,V, ...) \, ,
    \label{eq:update}
\end{align}
where $T$ denotes the unknown tensor object, $V$ denotes the (nuclear) interaction, and the function $f(\cdot)$ encodes the specifics of the underlying few- or many-body framework.
The ellipsis indicates the possible presence of additional (method-specific) tensors in a given framework.

We start from a factorized representation of the many-body tensor
\begin{align}
    T = \prod_{i=1}^m X^{(i)} \, ,
    \label{eq:tensdec}
\end{align}
where the objects $X^{(i)}$ define the factors of the decomposition (see, \eg{}, Ref.~\cite{Kolda2009} for a review on tensor decompositions).
Obtaining computational benefits from such a factorization requires the reformulation of the few- or many-body formalism, fully operating on the factors themselves instead of the initial (undecomposed) tensors.
Practically, this yields a new set of equations
\begin{align}
    X^{(1)} &= g_1 ( X^{(1)}, ..., X^{(m)},V, ...) \, , \notag \\
        &\,\,\, \vdots \\
    X^{(m)} &= g_m ( X^{(1)}, ..., X^{(m)},V, ...) \, , \notag
\end{align}
where the update functions $g_i$ depend on the chosen decomposition.
In addition, we can also employ a factorized form for the interaction $V$ using a potentially different tensor format.

\subsection{Tensor format}

The general strategy laid out here can be applied to different tensor formats.
In this work, we focus on the singular value decomposition (SVD) of an $N\times N$ matrix $M$, which we take to be real for simplicity,
\begin{align}
    M = L \Sigma R^\dagger \, ,
\end{align}
where $(\cdot)^\dagger$ denotes the Hermitian adjoint.
The diagonal matrix $\Sigma= \operatorname{diag}(s_1,...,s_N)$ contains the ordered set of nonnegative singular values $s_i$, and the left and right matrices of singular vectors, $L$ and $R$, are unitary.
By keeping only the leading $\RSVD$ singular values $s_1,...,s_{\RSVD}$ and the corresponding columns of the $L$ and $R$ matrices we obtain the truncated SVD of the matrix $M$ (indicated by the tilde)
\begin{align}
    \tilde M = \tilde L \tilde \Sigma \tilde R^\dagger \, ,
\end{align}
which provides a rank-$\RSVD$ approximation to the initial matrix.
In the following, we assume a factorized form for the two-body interaction
\begin{align}
    \tilde V = \tilde L_V \tilde \Sigma_V \tilde R_V^\dagger 
\end{align}
and similarly for the unknown many-body tensor\footnote{In the following, we suppress the tilde originally introduced to distinguish low-rank components $\tilde L_X,\tilde \Sigma_X, \tilde R_X$ from their full-rank counterparts $L_X,\Sigma_X,R_X$.}
\begin{align}
    \tilde T = \tilde L_T \tilde \Sigma_T \tilde R_T^\dagger \, .
    \label{eq:TFactors}
\end{align}
While many other matrix decompositions exist, \eg{}, eigenvalue and Cholesky decompositions, the SVD format is particularly versatile since it requires neither normality nor positive definiteness of the matrix.

In quantum chemistry, factorizations into even more tensors have been applied in the context of MBPT and CC calculations~\cite{Hohenstein2013,Schutski2017,Hohenstein2022}.
The employed \textit{tensor hypercontraction} (THC) formats are governed by a larger amount of decomposition factors, \ie{}, $m=5$ in Eq.~\eqref{eq:tensdec} as opposed to $m=3$ for the case of the SVD.
For a discussion of the THC format in the context of nuclear theory, see Refs.~\cite{Tich19THC,Tichai2019pre}.

\subsection{Minimization procedure}

\begin{figure}[t!]
    \centering
    \includegraphics[clip=,width=0.8\columnwidth]{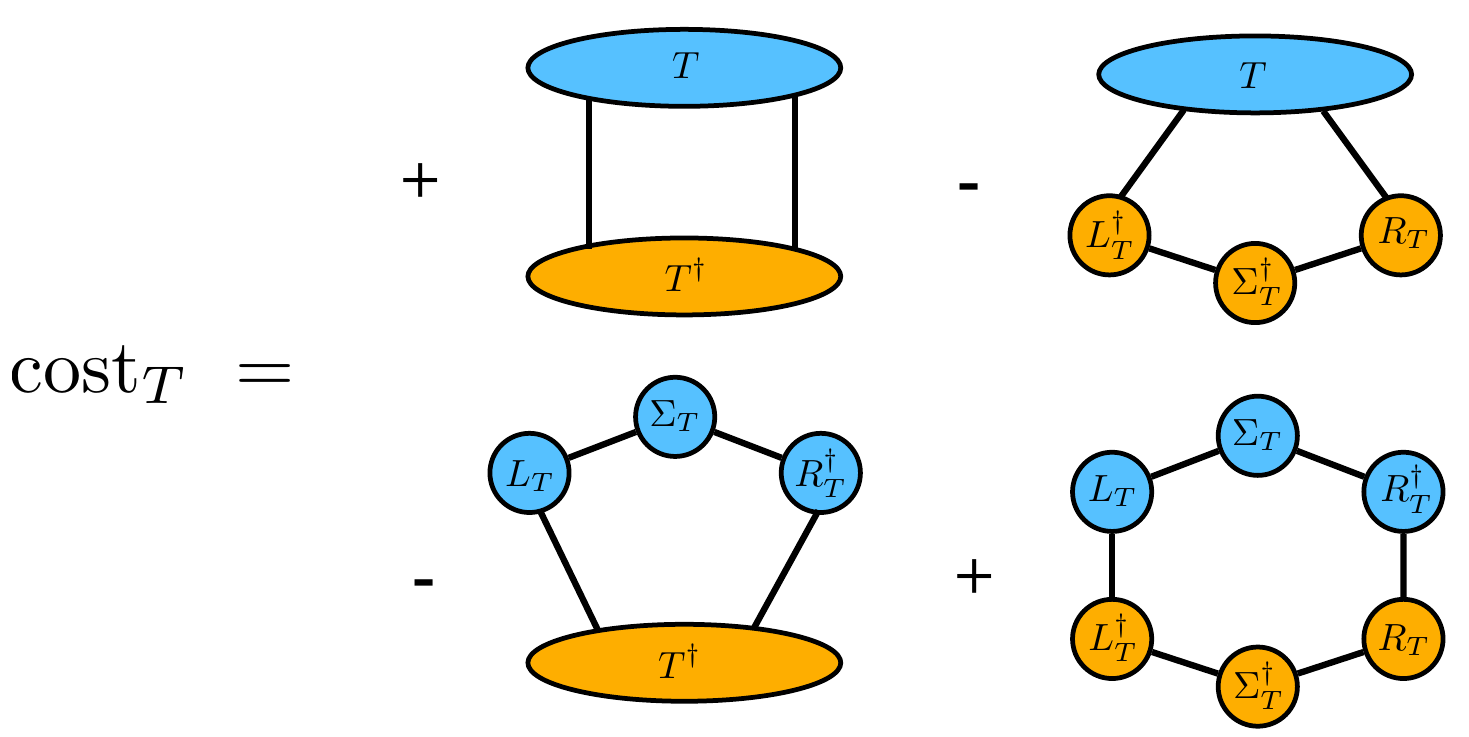}
    \caption{Diagrammatic representation of the cost function, Eq.~(\ref{eq:cost}).}
    \label{fig:cost}
\end{figure}

\begin{figure}[t!]
    \includegraphics[clip=,width=\columnwidth]{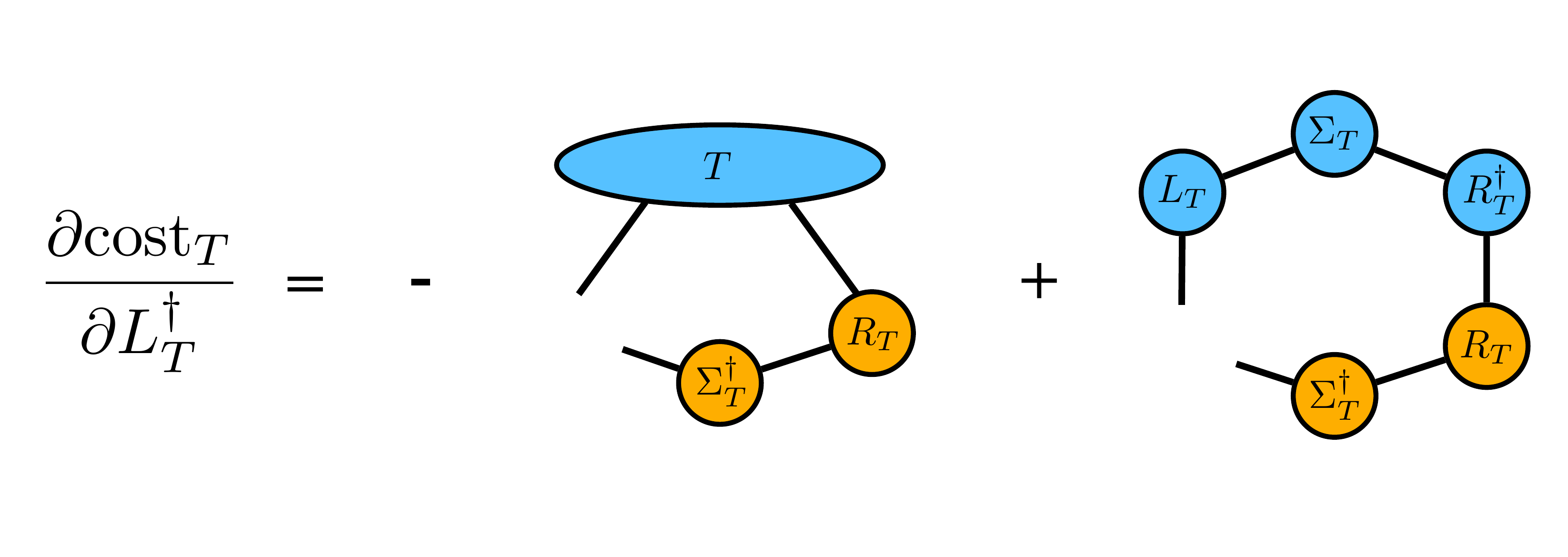}
    \caption{Derivative of the cost function with respect to $L_T^\dagger$.}
    \label{fig:deriv}
\end{figure}

We follow a least-square approach that minimizes the distance of the decomposed tensor to its original counterpart~\cite{Schutski2017}.
By introducing the error tensor $\Delta_T \equiv  T - \tilde T$ we define the cost function
\begin{align}
    \cost_T 
    &\equiv \Vert \Delta_T \Vert_\text{Fro.}^2 \, ,
\end{align}
where $\Vert \cdot \Vert_\text{Fro.}$ denotes the Frobenius norm,
\begin{align}
    \cost_T = \Tr [(T^\dagger - R_T \Sigma_T^\dagger L_T^\dagger)(T - L_T \Sigma_T R_T^\dagger)] \, .
    \label{eq:cost}
\end{align}
This can be diagrammatically represented as shown in Fig.~\ref{fig:cost}, where orange symbols indicate Hermitian adjoints of the blue symbols and connecting lines correspond to tensor contractions.
\noindent The factorized working equations are obtained by optimizing $\cost_T$, \ie{}, setting partial derivatives with respect to the decomposition factors to zero
\begin{align}
    \frac{\partial \cost_T}{\partial X} = 0 \, ,
    \label{eq:partder}
\end{align}
where $X \in \{L_T,\Sigma_T, R_T^\dagger\}$.
Because the function $\cost_T$ is real-valued and analytic,
\begin{align}
    \frac{\partial \cost_T}{\partial X} = \bigg( \frac{\partial \cost_T}{\partial X^\dagger} \bigg)^\dagger \, ,
\end{align}
derivatives with respect to $X$ and $X^\dagger$ are linearly dependent and only one set must be taken into account.
Diagrammatically, performing a derivative $\partial \cost_T/ \partial X$ corresponds to the removal of the corresponding tensor vertex $X$ in the tensor network (see Fig.~\ref{fig:deriv} for the derivative with respect to $X=L_T^\dagger$).
For the various decomposition factors we obtain from Eq.~\eqref{eq:partder} the following set of derivatives for SVD-based decompositions:%
\begin{subequations}
\begin{align}
    \frac{\partial \cost_T}{\partial L_T^\dagger} &=
    - f(T) R_T \Sigma_T^\dagger
    + L_T \Sigma_T R_T^{\dagger} R_T \Sigma_T^\dagger
    \, , \\
    \frac{\partial \cost_T}{\partial \Sigma_T^{\dagger}} &=
    - L_T^{\dagger}f(T)R_T + L_T^{\dagger}L_T\Sigma_T R_T^{\dagger}R_T
    \, , \\
    \frac{\partial \cost_T}{\partial R_T} &=
    - \Sigma_T^{\dagger} L_T^{\dagger} f(T)
    + \Sigma_T^{\dagger} L_T^{\dagger} L_T \Sigma_T R_T^{\dagger} \, ,
\end{align}
\label{eq:derivatives}%
\end{subequations}
where $f(T)$ encodes the (non-factorized) working equation, Eq.~(\ref{eq:update}). The specific example of the Lippmann-Schwinger equation will be discussed in Sec.~\ref{sec:lsls}.

\subsection{Master equations}

\begin{figure}[t!]
    \centering
    \includegraphics[clip=,width=0.75\columnwidth]{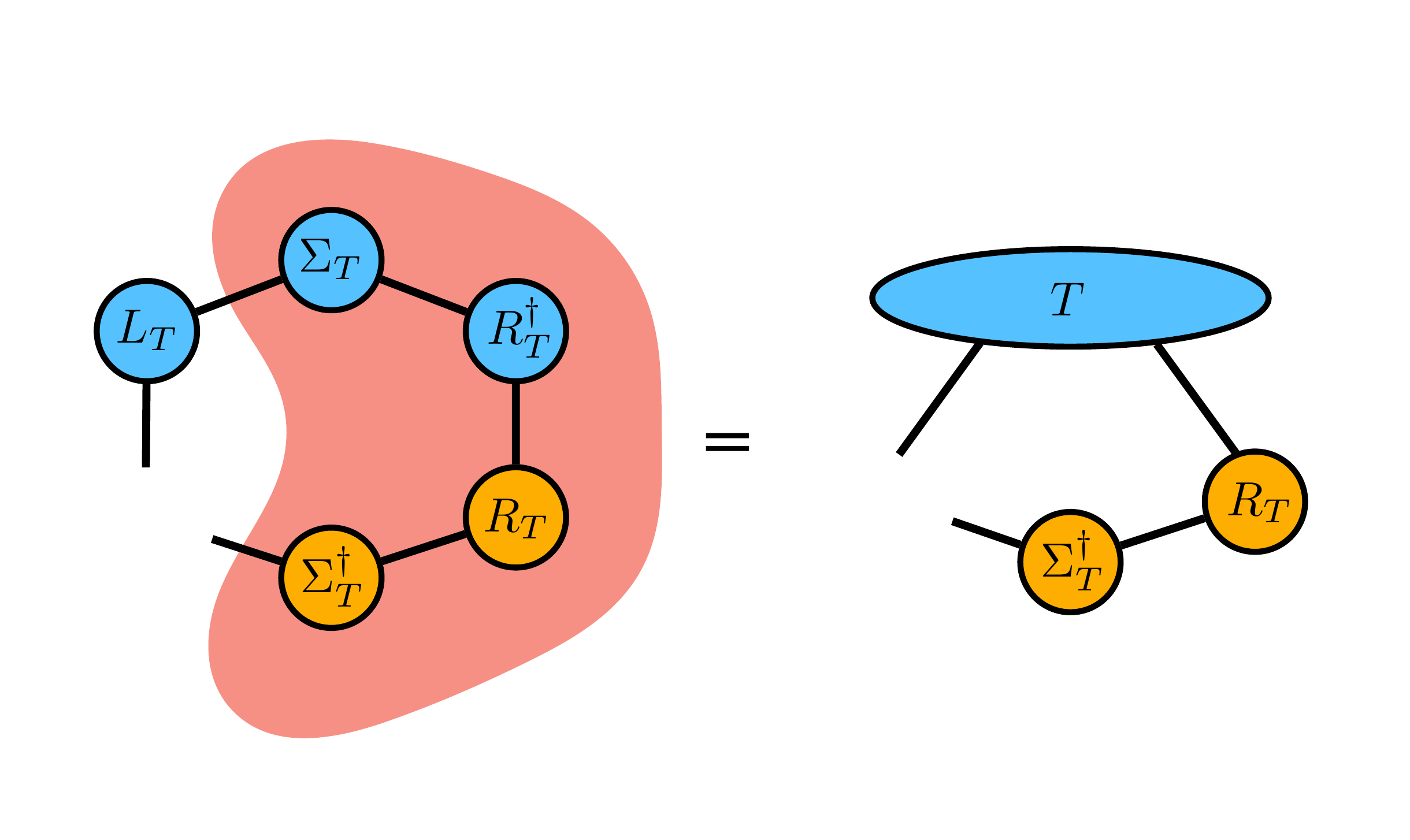}
    \caption{Diagrammatic representation of the derivative $\partial \cost_T/\partial L_T^\dagger = 0$. The colored area corresponds to the environment matrix (see text for details).}
    \label{fig:env}
\end{figure}
Since the derivative $\partial \cost_T/\partial X^\dagger$ is linear in $X$, all factors other than $X$ can be contracted in a so-called environment matrix $A_X$.
Figure~\ref{fig:env} shows the example of setting the derivative with respect to $L_T^{\dagger}$ to zero, where the contraction of the colored area gives the environment matrix $A_L$ associated with $L_T$.
We are left with the solution of a linear problem $X \cdot A_X = B_X$, where $B_X$ corresponds to the first terms in Eqs.~\eqref{eq:derivatives} and we note that the environment matrix can be a left and/or right factor.
Thus, the update step can be written as
\begin{align}
    X = B_X \cdot A^{-1}_X \, .
    \label{eq:update}
\end{align}
The explicit expressions for the environment matrices are%
\begin{subequations}
\begin{align}
    A_L &= 
    \Sigma_T R_T^\dagger R_T \Sigma_T^\dagger \, , \\
    A_{\Sigma_1}^{\text{}} &= 
    L_T^\dagger L_T \, , \\
    A_{\Sigma_2}^{\text{}} &= 
    R_T^\dagger R_T \, , \\
    A_R &= 
    \Sigma_T^\dagger L_T^\dagger L_T \Sigma_T \, .
\end{align}
\label{eq:env}%
\end{subequations}
In contrast to the left and right matrices the derivative $\partial \cost_T/\partial \Sigma_T^\dagger$ produces two environment matrices $A_{\Sigma_1}$ and $A_{\Sigma_2}$.
Note that the environment matrices are tensor-format specific and do not depend on the many-body approach itself (which is encoded in the tensors $B_X$).
Thus, the system of Eqs.~\eqref{eq:derivatives} constitutes a set of master equations governing any SVD-structured framework using algebraic equations.
Finally, the update steps for the different factor matrices are given by%
\begin{subequations}
\begin{align}
    L_T &= f(T) R_T \Sigma_T^\dagger A_L^{-1} \, , \\
    \Sigma_T &= A_{\Sigma_1}^{-1} L_T^\dagger f(T) R_T A_{\Sigma_2}^{-1} \, , \\
    R_T^\dagger &= A_R^{-1} \Sigma_T^\dagger L_T^\dagger f(T)\, ,
\end{align}
\label{eq:master}%
\end{subequations}
where all evaluations in the tensor-structured framework can be performed using efficient linear algebra operations.
In the following, we will refer to the update equations, Eqs.~\eqref{eq:master}, as the SVD-factorized least square (\tfls{}) equations.

\subsection{Explicit orthogonalization}

The solution of Eqs.~\eqref{eq:master} is not constrained to conserve unitarity of the left and right matrices, \ie{}, $L_T^\dagger L_T \neq 1 \neq R_T^\dagger R_T$.
Unitarity can be explicitly enforced  through an additional orthogonalization from a QR factorization,%
\begin{subequations}
\begin{align}
 L_T &= Q_L R_L \, , \\
 R_T &= Q_R R_R \, ,
\end{align}%
\end{subequations}
where $Q_R, Q_L$ are unitary matrices and $R_L, R_R$ are upper triangular matrices.
Modified factor matrices are obtained via%
\begin{subequations} 
\begin{align}
    \breve L_T & = Q_L \, , \\
    \breve  \Sigma_T & = R_L \Sigma_T R_R^\dagger \, , \\
    \breve  R_T & = Q_R \, , 
\end{align}%
\end{subequations}
such that the left and right matrices are manifestly unitary. 
To restore the diagonality of $\Sigma_T$ we perform an additional (untruncated) SVD
\begin{align}
    \breve \Sigma_T = \bar L_T \bar \Sigma_T \bar R^\dagger_T \,,
\end{align}
where the unitary matrices $\bar L_T$ and $\bar R_T$ are absorbed into the left and right matrices, respectively, and we obtain the final decomposition
\begin{subequations}
\begin{align}
    \check L_T &=  Q_L \bar L_T \, , \\
    \check \Sigma_T &=  \bar \Sigma_T \, , \\
    \check R^\dagger_T &=  \bar R^\dagger_T  Q_R^\dagger  \, .
\end{align}%
\end{subequations}
Practically, the orthogonalization is performed in each iteration step. Due to the small size of the corresponding matrices the computational overhead is negligible.

The restoration of a proper SVD format has the formal advantage of simplifying the master equations of the least-square approach.
Due to the diagonality of the $\Sigma_T$ matrix the environment matrices [Eqs.~\eqref{eq:env}] can be analytically inverted giving rise to the update step%
\begin{subequations}
\begin{align}
    L_T &= f(T) \check R_T \check \Sigma_T^{-1} \, , \\
    \Sigma_T &= \check L_T^\dagger f(T) \check R_T  \, , \\
    R_T^\dagger &= \check \Sigma_T^{-1} \check L_T^\dagger f(T)\, .
\end{align}
\label{eq:masterorth}%
\end{subequations}
The simplified update step in Eqs.~\eqref{eq:masterorth} reduces the numerical sensitivity to ill-conditioned environment matrices in the formation of their inverses.
In the following, we refer to the \tfls{} approach as the one with explicit orthogonalization, Eqs.~\eqref{eq:masterorth}.

\subsection{Computational advantages}

The computational benefit of factorization techniques comes from the decreased memory requirements and lower number of floating point operations in the tensor contractions, both of which depend on the dimension of the matrix objects.

The evaluation cost of a matrix-matrix product scales as $\contr = \scaling{N^3}$ for dense $N \times N$ matrices each with an associated memory cost $\stor= \scaling{N^2}$.
In presence of an SVD-factorized matrix the \tfls{} update equations induce an operation count of $\fcontr = \scaling{N^2 \RSVD}$ with memory cost $\fstor = \scaling{N \RSVD}$ for the factorization components.
The full scaling is recovered in the limit of an exact decomposition, which in the case of an SVD corresponds to keeping all $N$ singular values,%
\begin{subequations}
\begin{align}
    \lim_{\RSVD\rightarrow N} \fcontr &= \contr \, , \\
    \lim_{\RSVD\rightarrow N} \fstor &= \stor \, .
\end{align}%
\end{subequations}
We note that this statement is only true asymptotically, since the SVD induces a storage overhead in absence of a truncation. However, this will only affect the prefactor and not the scaling exponent.
In practice, obtaining benefits through factorizations relies on the low-rank properties of the tensors such as the input interaction, \ie{}, how accurate low-rank approximations are at $\RSVD \ll N$.

\section{Two-body scattering}
\label{sec:lsls}

\subsection{Lippmann-Schwinger equation}

\begin{figure}[t!]
\centering
\includegraphics[clip=,width=\columnwidth]{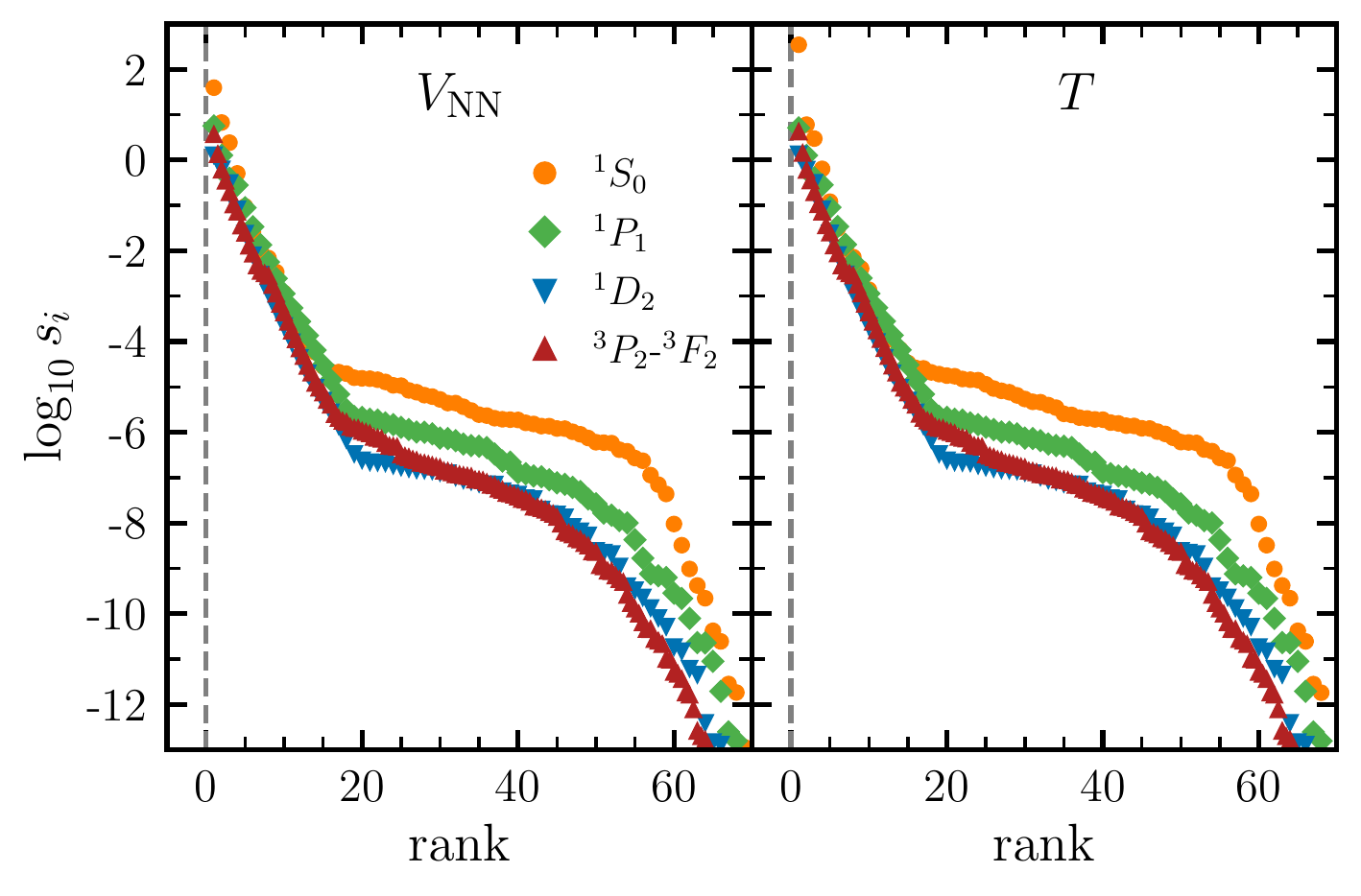}
\caption{Singular values (in \fm) for the $V_{\NN}$ and $T$ matrices in different partial-wave channels for the \nxlo{3} EMN~500 potential. For the coupled $^3P_2$-$^3F_2$ channel the rank is divided by two.}
\label{fig:svalsVT}
\end{figure}

\begin{figure*}[t!]
\centering
\includegraphics[clip=,width=0.75\textwidth]{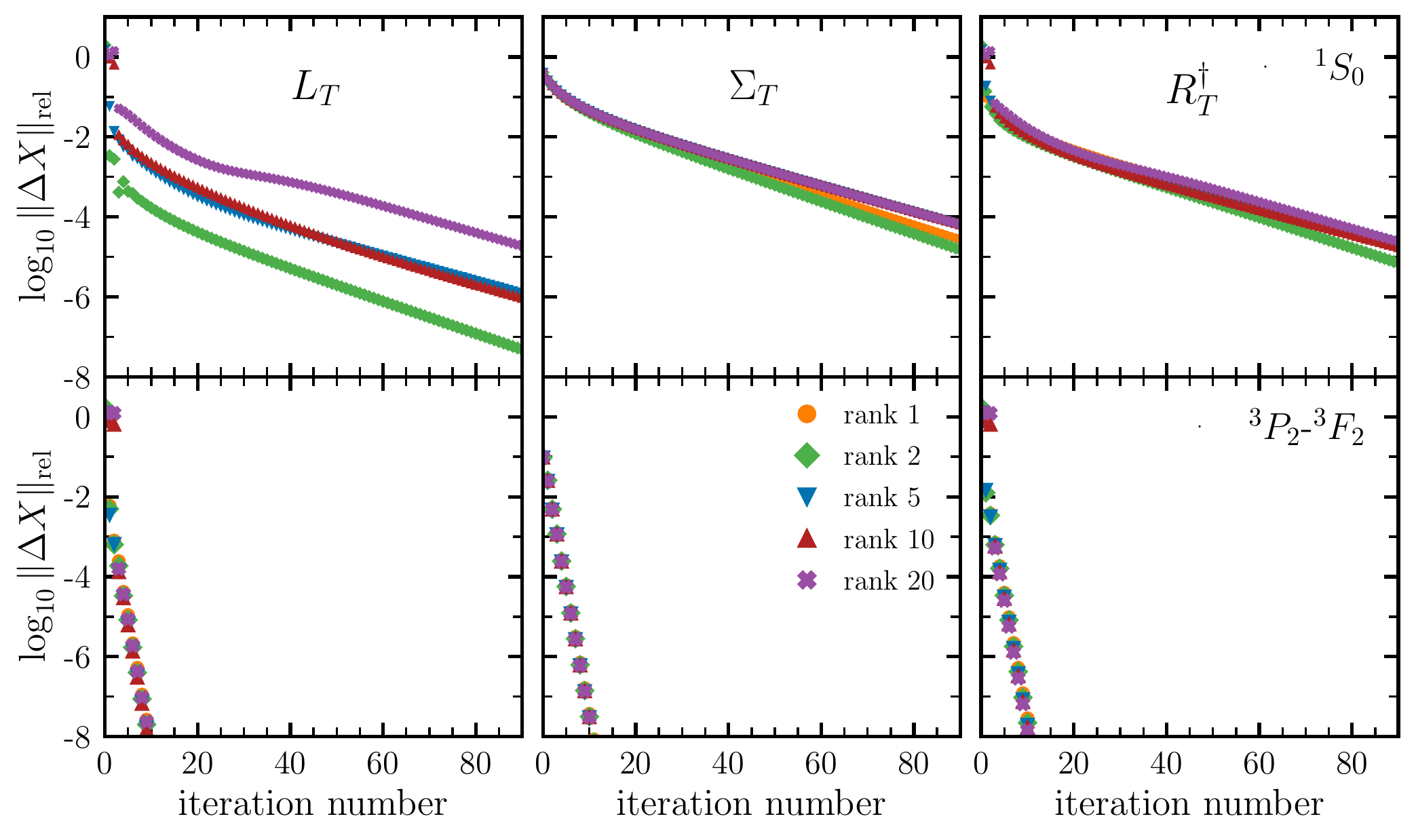}
\caption{Convergence of the decomposition factors of the $T$ matrix in the \tfls{} approach as function of iteration number in the $^1S_0$ channel (upper panels) and the coupled \coupledchannel{P}{F}{2} partial wave (lower panels) for the \nxlo{3} EMN~500 potential. Results are shown for different rank approximations.}
\label{fig:convergence}
\end{figure*}

In the following, we systematically apply the least-square approach to the \LippSchw{} equation 
\begin{align}
    T = V + V G_0 T \, ,
    \label{eq:lse}
\end{align}
where $G_0$ denotes the (diagonal) free Green's function and $T$ the $T$ matrix, which we take to be right-side half-on-shell.
Equation~\eqref{eq:lse} is of the general form of an algebraic update equation, Eq.~\eqref{eq:update}, with
\begin{align}
    f(T,V,G_0) = V+ V G_0 T \, .
\end{align}
For the initialization of the $T$ matrix factors the first-order Born approximation $T^{(0)} = V$ is employed leading to $X_T^{(0)} = X_V $ for $X\in \{L, \Sigma,R^\dagger\}$, thus requiring the same target rank for the $T$ matrix and the potential.

The two-nucleon (\NN{}) potential (and the $T$ matrix) are represented in a partial-wave basis
\begin{equation}
\la k \, (\lang S) J T M_T | V_\NN | k' \, (\langpr S) J T M_T \ra\,,
\end{equation}
with the final and initial orbital angular momenta  $\lang$ and $\langpr$, the two-body spin $S$, the two-body total angular momentum $J$, the two-body isospin $T$ with projection $M_T$, and the absolute values of the outgoing and incoming relative momenta $k$ and $k'$. In each partial-wave channel, the \NN{} potential is represented using $N=100$ momentum mesh points up to $k_{\text{max}}=k_{\text{max}}'=6.0\,\fm^{-1}$.
Similarly, the \LippSchw{} equation is solved in a partial-wave-decomposed form (with $\hbar^2/m=1$),
\begin{align}
    &\la k \alpha | T(E=k'^2)| k'  \alpha' \ra =
    \la k \alpha | V | k'  \alpha' \ra 
    \notag \\[2mm]
    &+ \frac{2}{\pi} \sum_{\alpha''} \mathcal{P} \int_0^\infty dq \, q^2 \,
    \frac{ \la k \alpha| V | q \alpha'' \ra \la q \alpha'' | T(E=k'^2) | k' \alpha' \ra }{k'^2 - q^2} \, ,
    \label{eq:lspw}
\end{align}
where $\alpha, \alpha', \alpha''$ are collective labels for the partial-wave quantum numbers.

\subsection{\emph{A priori} decomposition analysis}

Before turning to the study of the \tfls{} approach, we begin with a study of the low-rank properties of the $T$ matrix obtained from direct-inversion techniques of the Lippmann-Schwinger equation.
In the following, we employ the \nxlo{3} \NN{} potential from
Entem, Machleidt, and Nosyk (EMN) with a cutoff $\Lambda=500 \, \MeV$~\cite{Ente17EMn4lo}. Note that the singular value spectrum
is qualitatively similar for different orders, different cutoffs,
and with similarity renormalization group (SRG) evolution~\cite{Tich21SVDNN,Zhu21SVD}.

Figure~\ref{fig:svalsVT} shows a comparison of the singular spectrum of the initial \NN{} potential and the final right-side half-on-shell $T$ matrix for different partial-wave channels.
Similarly to Ref.~\cite{Tich21SVDNN}, we divide the rank in the coupled channel by a factor of two to be able to compare channels with different matrix dimensions.
It is evident that the initial low-rank properties directly propagate to the $T$ matrix and the $T$ matrix itself is dominated by very few components in the SVD expansion.
For the $T$ matrix in the \uncoupledchannel{S}{0} channel there is a strong enhancement of the leading singular value $s_1$ by a factor of ten going from $V$ to the $T$ matrix due to the large scattering length.

\subsection{Numerical convergence of the\\ least-square approach}

\begin{figure*}[t!]
    \centering
    \includegraphics[clip=,width=1.0\textwidth]{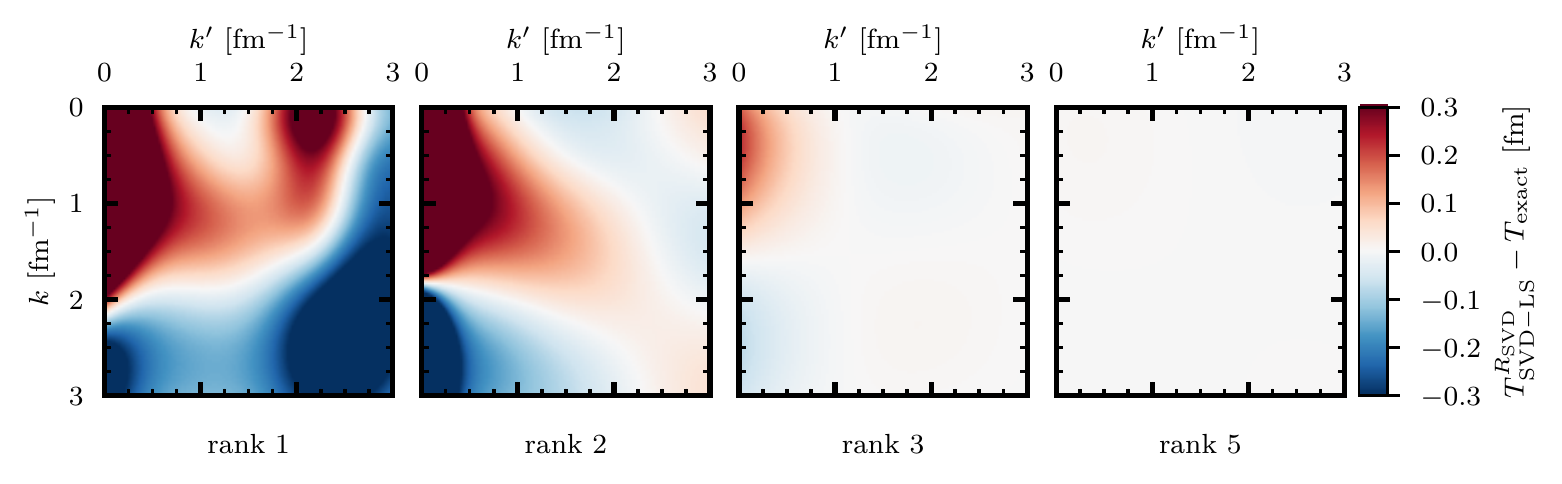}
    \caption{Absolute difference of low-rank $T$ matrices obtained from the least-square approach compared to the exact solution of the \LippSchw{} equation via direct inversion. Results are shown for the \nxlo{3} EMN~500 potential in the $^1S_0$ channel.}
    \label{fig:Tmatsvd}
\end{figure*}

The self-consistent solution of the \tfls{} equations is obtained by consecutive updates of $L_T, \Sigma_T, R_T^\dagger$ while keeping all other factors fixed.
Convergence is gauged by the relative norm of the difference between consecutive factor matrices,
\begin{align}
    \Vert \Delta X \Vert_\text{rel} \equiv \frac{\Vert X^{(n+1)} - X^{(n)} \Vert}{\Vert X^{(n)} \Vert} \, ,
\end{align}
where the superscript indicates the iteration number.

Figure~\ref{fig:convergence} shows the rate of convergence for the decomposition factors as a function of iteration number for the \uncoupledchannel{S}{0} and the coupled \coupledchannel{P}{F}{2} partial wave.
Clearly the large-scattering-length \uncoupledchannel{S}{0} channel requires a significantly larger number of iterations compared to the weaker \coupledchannel{P}{F}{2} channel.
Moreover, the convergence for the \uncoupledchannel{S}{0} channel strongly depends on the initial rank of the potential, with a significant increase of iterations needed until convergence beyond $\RSVD \approx 10$.
We attribute this to numerical instabilities in the inversion of the environment matrices due to small singular values at higher rank (\eg{}, $s_{20} \lesssim 10^{-5}$). However, these high-rank components are not important for an accurate reproduction of the NN $T$ matrix.
For the \uncoupledchannel{S}{0} channel the rate of convergence of the decomposition factors $\Sigma_T$ and $R_T^\dagger$ is slower than for $L_T$, in particular at low rank $\RSVD \lesssim 3$.
This behavior is likely related to the use of the right-side half-on-shell $T$ matrix, $\langle k| T(E=k'^2)|k'\rangle$, so that the iteration is sensitive to the energy dependence of the free Green's function $G_0(E=k'^2)$ around $E=k'^2$.

The partial-wave dependence of the rate of convergence can be understood from an analysis of the integral kernel $K(E) = V G_0(E)$ that enters the \LippSchw{} equation.
In an iterative approach, the final $T$ matrix is obtained as the infinite Born series
\begin{align}
    T(E) = V + \sum_{k=1}^\infty \, \Bigl[ VG_0(E) \Bigr]^k V \, ,
\end{align}
so that $K(E)$ drives the numerical stability of the least-square approach.
This can be quantified in terms of the spectral radius of the integral kernel
\begin{align}
    \rho_\text{spec}\bigl(K(E)\bigr) = \max_i |\lambda_i| \, ,
\end{align}
where $\lambda_i$ are the eigenvalues of $K(E)$.
As the Born series constitutes a geometric series, eigenvalues $|\lambda_i| \geqslant 1$ will prevent an iterative approach from converging.
While eigenvalues $|\lambda_i|$ close to unity will not necessarily prevent convergence, they induce a much lower rate of convergence in practice.
For the deuteron \coupledchannel{S}{D}{1} channel, the presence of the bound state leads to an eigenvalue $|\lambda_i|=1$ at the deuteron binding energy $E = -2.2245 \, \MeV$, so that we do not present results for the deuteron channel.

This problem of non-convergence in an iterative approach can be circumvented using direct inversion techniques, which is easily possible for two-body scattering because of the small matrix dimension. However, already in the three-body sector one relies on iterative schemes and thus naturally encounters diverging Born series (see, \eg, Ref.~\cite{Miller2022ndscattering}).
This can be resolved by employing Pad\'e resummation techniques on the individual terms of the Born series, which enables a robust extraction of scattering observables~\cite{Gloe83QMFewbod}.

\begin{figure*}[t!]
    \centering
    \includegraphics[clip=,width=1.0\textwidth]{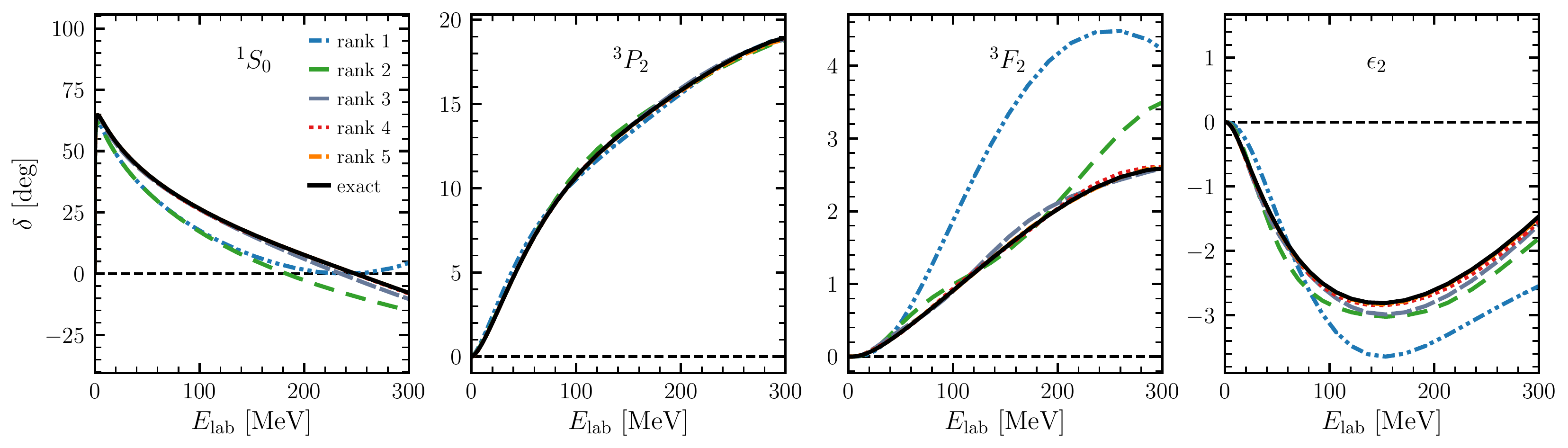}
    \caption{Two-nucleon phase shifts as a function of laboratory energy in the \uncoupledchannel{S}{0} and \coupledchannel{P}{F}{2} partial waves based on the low-rank $T$ matrices obtained from the least-square approach and in comparison to the exact $T$ matrix. Results are shown for the \nxlo{3} EMN~500 potential.}
    \label{fig:phaseshift}
\end{figure*}

\subsection{Diagnostic of the low-rank $T$ matrix solution}

We continue our analysis with the characterization of the solution of the \tfls{} approach.
The exact $T_\text{exact}$ matrix is obtained from the full-rank, $\RSVD=N$, interaction and by solving the \LippSchw{} equation via direct inversion~\cite{Gloe83QMFewbod}.
We compare the low-rank $T$ matrix from the least-square approach, $T^{\RSVD}_\text{\tfls} = L_T \Sigma_T R_T^\dagger$, to the exact $T$ matrix.
As error measure we study in the following absolute and relative differences of the matrix object,
\begin{subequations}%
\begin{align}
    \Vert \Delta T \Vert_\text{abs} &= 
    \Vert T- T_\text{exact} \Vert \, , \\
    \Vert \Delta T \Vert_\text{rel} &= 
    \frac{\Vert T- T_\text{exact} \Vert}{\Vert T_\text{exact} \Vert} \, .
\end{align}%
\end{subequations}
A matrix plot of the low-rank $T$ matrix compared to the exact solution is provided in Fig.~\ref{fig:Tmatsvd}.
At rank $\RSVD=1$ and $2$ we observe a sizeable difference from the full-rank $T$ matrix, in particular in the low-momentum regime $k,k' \lesssim 2\, \fm^{-1}$.
Once the rank is increased, deviations decrease systematically yielding only minor differences for $\RSVD=3$ and excellent agreement at $\RSVD=5$.

\begin{table}[t!]
\caption{Comparison of low-rank $T$ matrices and singular values obtained from the least-square approach to the exact $T$ matrix from direct inversion and to the exact singular values at full rank, $\RSVD=100$. All dimensionful quantities are in fm. As in Fig.~\ref{fig:Tmatsvd}, results are given for the \nxlo{3} EMN~500 potential in the $^1S_0$ channel.}
    \centering
    \renewcommand*{\arraystretch}{1.4}
    \begin{tabular*}{\columnwidth}{c@{\extracolsep{\fill}} ccc ccccc}
    \hline \hline
    \RSVD & $\Vert \Delta T \Vert_\text{rel}$ & $\Vert \Delta T \Vert_\text{abs}$ & $s_1$ & $s_2$ & $s_3$ & $s_4$ & $s_5$ \\
    \hline
    1 & 0.11 & 38.36 & 314.02 & $-$ & $-$ & $-$ & $-$ \\
    2 & 0.16 & 54.31 & 295.45 & 6.18 & $-$ & $-$ & $-$ \\
    3 & 0.011 & 3.86 & 345.19 & 6.06 & 2.95 & $-$ & $-$ \\
    4 & $9.8\times 10^{-4}$ & 0.34 & 348.60 & 6.05 & 2.97 & 0.64 & $-$ \\
    5 & $2.2 \times 10^{-4}$ & $7.7\times 10^{-2}$ & 348.63 & 6.04 & 2.96 & 0.64 & 0.12 \\
    10 & $2.5 \times 10^{-6}$ & $8.8 \times 10^{-4}$ & 348.67 & 6.04 & 2.97 & 0.64 & 0.12 \\
    20 & $1.5\times10^{-7}$ & $5.1 \times 10^{-5}$ & 348.67 & 6.04 & 2.97 & 0.64 & 0.12 \\
    \hline 
    100 & $-$ & $-$ & 348.67 & 6.04 & 2.97 & 0.64 & 0.12 \\
    \hline \hline
    \end{tabular*}
    \label{tab:svdvt}
\end{table}

Using an initial \NN{} interaction, there are two ways to obtain the low-rank $T$ matrix:
\begin{itemize}
    \item[i)] \textit{Decompose and reconstruct:} Perform a low-rank approximation for the initial potential and use the truncated potential to obtain the low-rank $T$ matrix from direct inversion. In this case, the $T$ matrix factors are obtained from an explicit SVD of the resulting $T$ matrix.
    \item[ii)] \textit{Least-square approach:} Perform a low-rank approximation for the initial potential and use its decomposition factors as input for the least-square approach described in Sec.~\ref{sec:lse}. The final low-rank $T$ matrix is then given by the converged factors after the least-square iteration.
\end{itemize}
At fixed rank $\RSVD$, both strategies yield an equivalent final solution, up to unitary transformations among the left and right matrices due to the re-orthogonalization.

Table~\ref{tab:svdvt} shows the quality of low-rank $T$ matrices compared to the exact results. From rank $\RSVD=1$ to $2$ there is a slight increase in relative error, but for larger ranks the relative error systematically decreases as the rank is increased.
The leading singular values of the $T$ matrix do not remain constant as the rank is increased due to the nonlinear dependence between the singular values of $V$ and $T$. At rank $\RSVD=10$ the singular values stabilize and agree with the exact singular values at full rank, $\RSVD=100$.

\subsection{Two-nucleon phase shifts}
\label{sec:phaseshift}

We finally turn to the description of two-nucleon phase shifts based on low-rank $T$ matrices.
The phase shifts are obtained from the least-square $T$ matrix using the converged factor matrices $T = L_T \Sigma_T R^\dagger_T$.
Figure~\ref{fig:phaseshift} shows the phase shifts for the $^1S_0$ and the coupled \coupledchannel{P}{F}{2} partial waves using low-rank $T$ matrices with $\RSVD=1,\ldots,5$.
At very low rank, the phase shifts significantly deviate from the exact results. This is most pronounced in the $^3F_2$ partial wave and for the mixing angle $\epsilon_2$, while the $^3P_2$ channel shows very little sensitivity to the rank.
The enhanced sensitivity in the mixing angle has also been observed for the deuteron channel in Ref.~\cite{Tich21SVDNN}.
The quality of the approximation is systematically improved in all channels and at rank $\RSVD=5$ the phase shifts are excellently reproduced up to laboratory energies $E_\text{lab} \lesssim 300 \, \MeV$.
Since the low-rank $T$ matrices were already shown to reproduce the exact $T$ matrix very well (see Fig.~\ref{fig:Tmatsvd}), it is clear that derived quantities yield a similarly good approximation error.

\section{Conclusions and outlook}
\label{sec:outlook}

In this work, we have presented a new strategy to solve algebraic equations from a factorization ansatz.
Following a least-square approach, the master equations are derived from a stationarity condition of the cost function based on the error tensor.
We have derived a set of update equations for the individual factor matrices. This strategy is general and can be used to exploit tensor factorization techniques in few- and many-body calculations.
Moreover, the \tfls{} equations can be adapted to other tensor formats as previously shown in quantum chemistry applications~\cite{Schutski2017}.
The feasibility of the least-square approach is practically demonstrated for the Lippmann-Schwinger equation.
By employing an SVD form for the potential, a factorized form for the $T$ matrix was obtained.
Using an additional explicit orthogonalization during the self-consistent iterations enables the recovery of a proper SVD format for the $T$ matrix itself.
The right-side half-on-shell $T$ matrix and two-nucleon phase-shifts are in excellent agreement with full-rank calculations, reflecting the low-rank structures of chiral interactions~\cite{Tich21SVDNN,Zhu21SVD} and their propagation to the associated $T$ matrices.
We note, however, that the transformation to single-particle bases may lower the efficacy of the SVD approximation of nuclear potentials in the many-body calculations due to the coupling with center-of-mass degrees of freedom as discussed in Ref.~\cite{Zhu21SVD}.

While we have discussed computational benefits in terms of abstract scaling laws, their demonstration in our two-body application is difficult due to the very short runtimes of the order of hundreds of miliseconds.
Future applications to larger-scale simulations will allow for more systematic studies of the underlying computational benefits.

Our work also suggests extensions to nuclear few- and many-body calculations.
An interesting application will be to explore the least-square approach in nonperturbative many-body methods, such as CC or IMSRG calculations of medium-mass nuclei. These have the necessary algebraic equations for the coupled-cluster amplitudes or the evolution operator, respectively, which could benefit from factorization methods through the least-square approach.

\acknowledgments

This work was supported in part by the European Research Council (ERC) under the European Union's Horizon 2020 research and innovation programme (Grant Agreement No.~101020842), the Deutsche  Forschungsgemeinschaft  (DFG,  German Research Foundation) -- Project-ID 279384907 -- SFB 1245, the Helmholtz Forschungsakademie Hessen für FAIR (HFHF), and by the BMBF
Contract No.~05P21RDFNB.
Computations were in part performed with an allocation of computing resources at the Jülich Supercomputing Center.

\bibliography{strongint}

\end{document}